\documentclass[conference]{IEEEtran}
\usepackage{graphicx, epstopdf}
\usepackage{amssymb, amsmath}
\usepackage{commath}
\usepackage{setspace}
\usepackage{acronym}
\usepackage{mathrsfs}
\usepackage{ifthen}
\usepackage{textcomp}
\usepackage{float}
\usepackage{subfigure}
\usepackage{color}
\usepackage{cite}
\usepackage{amsthm}
\usepackage[version=3]{mhchem}

\IEEEoverridecommandlockouts

\begin{document}
\bstctlcite{ICC09_Ref2:BSTcontrol}

\title{A Novel Molecular Communication System\\ Using Acids, Bases and Hydrogen Ions}

\author{ \IEEEauthorblockN{Nariman~Farsad,~\IEEEmembership{Member,~IEEE,}
											and~Andrea~Goldsmith,~\IEEEmembership{Fellow,~IEEE,}}

\IEEEauthorblockA{Department of Electrical Engineering, Stanford University, CA, USA.
}
}
\maketitle

\begin{abstract}
Concentration modulation, whereby information is encoded in the concentration level of chemicals, is considered. One of the main challenges with such systems is the limited control the transmitter has on the concentration level at the receiver. For example, concentration cannot be directly decreased by the transmitter, and the decrease in concentration over time occurs solely due to transport mechanisms such as diffusion. This can result in inter-symbol interference (ISI), which can have degrading  effects on performance.  In this work, a new and novel scheme is proposed that uses the transmission of acids, bases, and the concentration of hydrogen ions for carrying information. By employing this technique, the concentration of hydrogen ions at the receiver can be both increased and decreased through the sender's transmissions. This enables novel ISI mitigation schemes as well as the possibility to form a wider array of signal patterns at the receiver.
\end{abstract}


\section{Introduction}
\IEEEPARstart{M}{olecular} communication is a new and emerging field, where information is conveyed through chemical signals \cite{Farsard_arXiv14}.  In this paradigm, the transmitter releases tiny particles, where information is modulated onto the chemical properties of these particles.  The particles then propagate through the medium until they arrive at the receiver, where the chemical signal is demodulated and the information is decoded. Although there are many different forms of transport in molecular communication (e.g. active transport \cite{far11Bionet}), the most common propagation mechanism considered in the literature is diffusion-based transport with and without flow between the transmitter and receiver \cite{Farsard_arXiv14}. 

Different modulation schemes have been proposed for molecular communication including: concentration modulation \cite{mah10}, type-based modulation \cite{kim13}, and timing of release modulation \cite{far15GLOBCOM}; a summary of some of these techniques can be found in \cite{kur11}. Between these modulation schemes, concentration modulation has received the most attention in the literature, since it can be detected with relative ease. For example, in \cite{far13}, a simple and inexpensive over-the-air tabletop demonstrator was developed for a concentration-modulated molecular diffusion channel with flow.

One of the issues with concentration-modulated diffusion-based molecular communication is that the concentration at the receiver increases with consecutive transmissions and only decreases as the particles diffuse away. Since diffusion can be a slow process, this will result in significant inter-symbol interference (ISI) \cite{kil13}.  This is evident in the long tails associated with the system's impulse response  \cite{far15GLOBCOM}. Some previous works have proposed ISI mitigation techniques, such as the use of enzymes for destroying the chemicals that remain in the channel from previous transmissions \cite{noe14}. However, in these techniques the transmitter does not have control over the chemical reactions in the channel, nor can it actively decrease the concentration at the receiver.


In this work, a new and novel signaling scheme using acids, bases, and the concentration of hydrogen ions is proposed. In this scheme, the transmitter can release either a strong acid or a strong base. The strong acids and bases dissociate almost completely in aqueous solutions (i.e. solutions where water is the solvent) to form hydrogen ions and hydroxide ions, respectively \cite{principleChem-Book09}. Moreover, if used in low concentrations, strong acids and bases are not corrosive or destructive to the transmitter and receiver. Because of the water autoionization reaction, which will be explained in detail in this paper, concentration of hydrogen ions and hydroxide ions are almost always non-zero, and the product of concentration of both ions is a constant. Therefore, an increase in the concentration of one species results in a proportional decrease in the concentration of the other species. The received signal at the destination is obtained by measuring the pH level, where pH is the negative log of the concentration of hydrogen ions. 
 
There are multiple benefits to using this scheme. First, for detection at the receiver, pH sensors are available at micro-scales and macro-scales, which makes this technique practical. Second, the concentration of hydrogen ions at the receiver can be directly increased {\em and decreased} by the transmitter. Third, a wider array of signal patterns can be formed at the receiver using this extra degree of freedom, which could then be used in different applications such as to generate control signals for synthetic biological devices, or as non-binary modulation schemes. For example, it was recently shown that pH signals could be used to control the motion direction of bacteria \cite{zhu15}. 
In this work we focus on ISI mitigation, and show that the proposed system could significantly reduce the ISI.

The rest of this paper is organized as follows. In Section \ref{sec:acidBase} the chemical reactions between the strong acids, strong bases, and hydrogen ions are presented. Then, in Section \ref{sec:sysModel} the system model for this channel is presented. The system response to an impulse of acid or base is studied in Section \ref{sec:sysResp}. An ISI mitigation technique is proposed in Section \ref{sec:isiMiti}, through consecutive transmission of an acid impulse followed by a base impulse. The paper ends with concluding remarks and future work in Section \ref{sec:conclusion}. 

\section{Acids, Bases and pH Scale}
\label{sec:acidBase}
In an aqueous environment, one important process is the autoionozation of water \cite{principleChem-Book09}. In this process, two water molecules are combined to generate a hydronium ion (\ce{H3O+})and a hydroxide ion (\ce{OH-}). The equation for this reaction is shown below:
\begin{align}
\label{eqc:h2oautoion}
	\ce{2H2O <=> H3O+(aq) + OH^-(aq)}.
\end{align}
Note that alternatively, in some texts, this reaction equation is simplified and written as
\begin{align}
\label{eqc:h2oautoion2}
\ce{H2O <=>   H+(aq) + OH^-(aq)}.
\end{align}
In this paper, the two equations and hence ions \ce{H+} (i.e. hydrogen ion) and \ce{H3O+} are used interchangeable. The equilibrium constant for the autoionozation reaction is given by  \cite{principleChem-Book09}
\begin{align}
\label{eq:autoIonEuilb}
k_w=\ce{[H+]}\ce{[OH^-]}, 
\end{align}
where the notation $[.]$ is used to indicate the concentration of ions and molecules. For water at 25$^\circ$C, $k_w=10^{-14}$. For pure water, since there are no other sources for ion formation, the concentration of both ions are equal and therefore we have
\begin{align}
\ce{[H+]}=\ce{[OH^-]} =10^{-7} \text{~M},
\end{align}
where M, called molar, is units of concentration and represents the number of moles per every liter of solution (i.e. mol/L). The molar concentration of a chemical can be related to the number of molecules of that chemical using the relation $M = \frac{N}{n_aL}$, where $N$ is the number of molecules, $n_a$ is Avogadro's number, and $L$ is the volume of solution. Therefore, the number of ions in every liter would be $10^{-7}\times6.022\times10^{23}=6.022\times10^{16}$, where the second number is Avogadro's constant.

The pH measure is the negative log scale of the concentration of hydrogen ions \ce{H+} (or \ce{H3O+}). Therefore, the pH of pure water at 25$^\circ$C is 
\begin{align}
\text{pH} =-\log (\ce{[H+]}) = 7.
\end{align}
Similarly, pOH represents the negative log scale of the concentration of hydroxide ions \ce{OH-}. At 25$^\circ$C  we have pH$+$pOH$=$p$k_w$=14 (p$k_w$ is the negative log of $k_w)$. At  this temperature the solution is neutral if $\text{pH}=7$, acidic if $\text{pH}<7$ (i.e. the concentration of hydronium/hydrogen ions is more than the concentration of hydroxide ions), and basic if $\text{pH}>7$ (i.e. the concentration of hydroxide ions is more than the concentration of hydronium/hydrogen ions). 

According to the Br{\o}nsted-Lowry definition of acids and bases, an acid is a proton donor, while a base is a proton acceptor  \cite{principleChem-Book09}\footnote{Although we use the Br{\o}nsted-Lowry definition of acids and bases in this paper, there are also the Arrhenius and the Lewis definitions of acids and bases. The Br{\o}nsted-Lowry definition, which is used in most textbooks, lends itself better to our analysis.}. Notice that according to this definition, water is both an acid and a base, since in (\ref{eqc:h2oautoion}), one water molecule donates a proton while the other one accepts a proton. 

An acid or a base is called a strong acid or a strong base if it almost completely dissociates in aqueous solutions (i.e. solutions where water is the solvent). Note that a strong acid or base does not necessarily result in an extremely low or high pH value. If the strong acid or base is used in low concentrations, the pH value of the solution would be close to the neutral pH, and the solution would not be corrosive or destructive. The dissociation reactions for strong acids and bases can be represented by the following chemical equations
\begin{align}
\label{eqc:aciddis}
\ce{AH(aq) + H2O(l)} &\ce{->  H3O+(aq) + A^-(aq)}, \\
\label{eqc:basedis}
\ce{B(aq) + H2O(l)} &\ce{->  BH+(aq) + OH^-(aq)},
\end{align}	
where \ce{AH} is a strong acid and \ce{B} is a strong base. In this work, the acid-base interaction that combines the $\ce{A-}$ and $\ce{BH+}$ to form a salt is ignored, since it does not have an effect on the concentration of hydrogen ions and the pH. 

If the quantity of the strong acid or base is small compared to the quantity of water, the dissociation happens almost instantly \cite{don12}. Using (\ref{eqc:h2oautoion}) and (\ref{eqc:aciddis}) together, it can be seen that the  \ce{H+} ion concentration increases, and the  \ce{OH-} ion concentration decreases as a strong acid is injected in the water. Similarly, using (\ref{eqc:h2oautoion}) and (\ref{eqc:basedis}) together, it is observed that the concentration of \ce{OH-} ions increases and the \ce{H+} ion concentration decreases as a strong base is dropped in the water. For example, if $6.022 \times 10^{22}$ molecules of a strong acid are dropped in a one liter water container, at steady state, the concentration of hydrogen ions would be $[\ce{H+}]=0.1$ M, the concentration of hydroxide ions would be $[\ce{H+}]=10^{-13}$ M, and the pH of the solution would be 1.

\section{System Model}
\label{sec:sysModel}
\begin{figure}[t]
	\begin{center}
		\includegraphics[width=0.9\columnwidth,keepaspectratio]{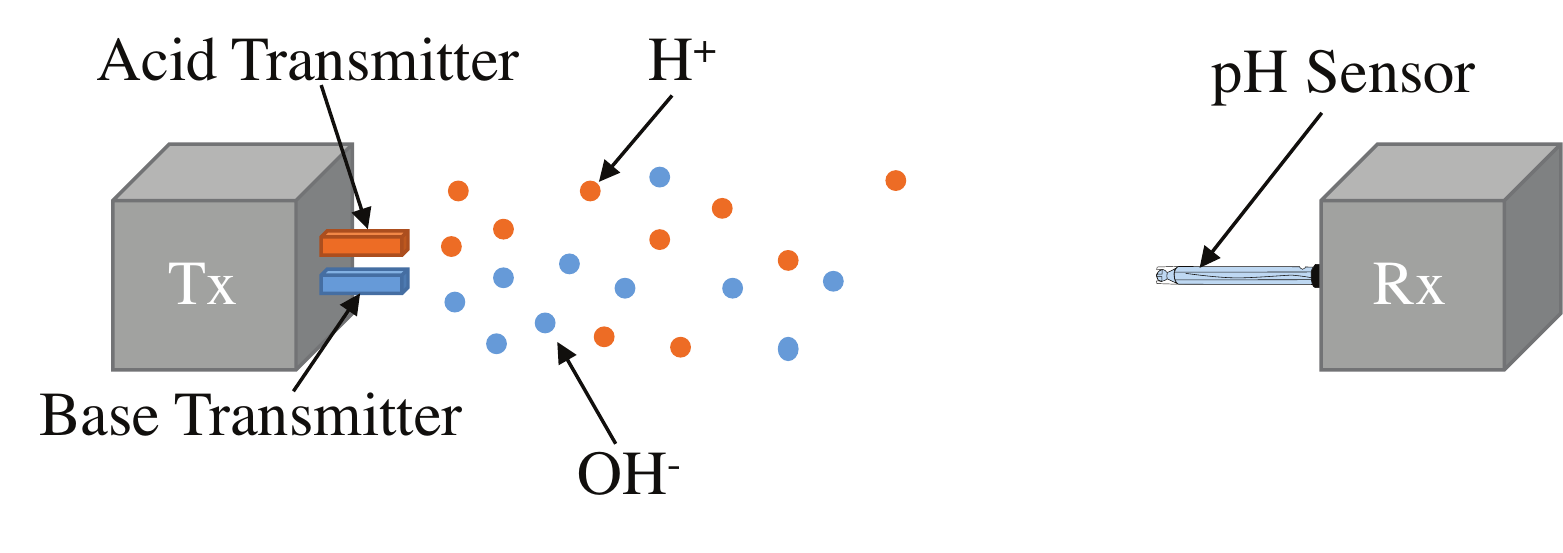}
	\end{center}
	\caption{\label{fig:pHcomm} Depiction of the proposed communication system.}
\end{figure}

The system that is  considered in this paper consists of a molecular communication transmitter that can release two different types of chemicals: a strong acid or a strong base. Note that a strong acid or base does not necessarily result in a very low or high pH value that could be destructive. If they are used in low concentrations, the pH levels could be kept closer to the neutral pH. Figure \ref{fig:pHcomm} depicts the proposed communication system. It is assumed that the transmitter can transmit both chemicals simultaneously in any concentration. It is also assumed that the distance between the ``nozzles'' that release the acid and base is small enough such that we can assume both chemicals are released from a single nozzle. The coordinate system that is used to study this communication channel is assumed to be centered at the tip of the nozzle, and the transmitter is a point source at this location.  

As was explained in the previous section, the strong acids and bases almost completely dissociate in water to form hydrogen ($\ce{H+}$) or hydroxide ($\ce{OH-}$) ions. This process happens very rapidly within a few microseconds \cite{don12}. Therefore, in this work it is assumed that transmitting a strong acid is equivalent to releasing hydrogen ions and transmitting a strong base is equivalent to releasing hydroxide ions. The released ions propagate  through convection-diffusion (or potentially pure diffusion) until they arrive at the receiver. The receiver would then use a pH meter to measure the pH level and detect the concentration of hydrogen ions. 

The only reaction involving the hydrogen and hydroxide ions in the channel is the water autoionization reaction given in (\ref{eqc:h2oautoion2}). Because the communication environment is inside a fluid, where the main solution is water, it can be assumed that there are always water molecules everywhere ready to dissociate. Therefore, reaction (\ref{eqc:h2oautoion2}) can be written as
\begin{align}
\label{eqc:autoiontonothing}
\ce{H+(aq) + OH^-(aq) <=>[k_f][k_r]} \varnothing
\end{align}
where $k_f$ and $k_r$ are the forward and reverse reaction rates, respectively.  At 25$^\circ$C the forward reaction rate for combination of $\ce{H+}$ and $\ce{OH-}$ ions is $k_f=1.4\times10^{11}$ 1/Ms \cite{physChemBook05}. The water dissociates into these ions at the rate $2.5\times10^{-5}$ 1/s  \cite{physChemBook05}. Since the molar concentration of water is 55.5 M, the reverse reaction rate is $k_r=2.5\times10^{-5} \times 55.5 = 1.4\times10^{-3}$ M/s, which represents the rate at which the two ions are generated \cite{physChemBook05}. Note that the forward reaction rate is much larger than the reverse reaction rate, which is the reason for very low ion concentrations of $10^{-7}$ M at equilibrium compared to 55.5 M of water.

 Let $C_{H}(\mathbf{x},t)$ and  $C_{OH}(\mathbf{x},t)$ represent the {\em average} spatiotemporal concentration of $\ce{H+}$ and $\ce{OH-}$ ions, respectively.  The {\em average} behavior of the transport system can be represented using a system of partial differential equations as
\begin{align}
\label{eq:reactDiffEqH}
\frac{\partial C_{H}}{\partial t} &=   D_{H} \nabla^2 C_{H}-\nabla.(\mathbf{v}C_{H})-k_f C_{H} C_{OH}+k_r  \\
\label{eq:reactDiffEqOH}
\frac{\partial C_{OH}}{\partial t} &=   D_{OH} \nabla^2 C_{OH}-\nabla.(\mathbf{v}C_{OH})-k_f C_{H} C_{OH}+k_r,
\end{align}
where $D_{H}$ and $D_{OH}$ are the diffusion coefficients of $\ce{H+}$ and $\ce{OH-}$ ions in water, and $\mathbf{v}$ is the velocity field. For pure diffusion the velocity field is assumed to be zero everywhere in the channel.
It is assumed that dimensions of the channel environment are much larger than the separation distance between the transmitter and  receiver. Therefore, infinite boundary conditions are assumed (i.e. $C_{H}(x=\infty,t)=C_{OH}(x=\infty,t)=0$).

The transmission process can be modeled by adjusting the initial conditions. Two scenarios are possible: only hydrogen ions (i.e. a strong acid) are released, or only hydroxide ions are released (i.e. a strong base). Note that if hydrogen ions and hydroxide ions (i.e. strong acid and base) are released simultaneously, because the forward reaction rate $k_f$ is very large, they immediately combine and neutralize to form water molecules. Therefore, the net effect would be one of the two scenarios mentioned earlier depending on the amount of hydrogen and hydroxide ions that were released. The two transmission processes can be represented by initial conditions
\begin{align}
\label{eq:initCondH}
\text{Tx releases $\ce{H+}$} &
\begin{cases}
C_{H}(\mathbf{x},t=0)=N_{H}\delta(\mathbf{x})+C_{H}^{\text{init}}(\mathbf{x})\\
C_{OH}(\mathbf{x},t=0)=C_{OH}^{\text{init}}(\mathbf{x})
\end{cases}\\
\label{eq:initCondOH}
	\text{Tx releases $\ce{OH-}$} &
	\begin{cases}
	C_{H}(\mathbf{x},t=0)=C_{H}^{\text{init}}(\mathbf{x}) \\
	C_{OH}(\mathbf{x},t=0)=N_{OH}\delta(\mathbf{x})+C_{OH}^{\text{init}}(\mathbf{x})
	\end{cases}
\end{align}
where the $N_{H}$ and $N_{OH}$ are the moles of hydrogen and hydroxide ions released by the transmitter, $C_{H}^{\text{init}}$ and $C_{OH}^{\text{init}}$ are the initial concentration profiles of each ion in the channel, and $\delta(.)$ is the vector form of the Dirac delta function. For example, if $\mathbf{x}$ represents a three dimensional space, then $\delta(\mathbf{x}) = \delta(x)\delta(y)\delta(z)$  where $x, y$ and $z$ are each axis of the coordinate system. For the very first transmission, we assume that $C_{H}^{\text{init}}=C_{OH}^{\text{init}}=10^{-7}$ (i.e. only water is present in the channel). This assumption is made to simplify our numerical analysis, and can be relaxed without affecting the model.

Unfortunately, an analytical solution to the system of nonlinear partial differential equations in (\ref{eq:reactDiffEqH})--(\ref{eq:reactDiffEqOH}) with the given boundary and initial conditions does not exist. Therefore, in the next section, the solutions for the two initial conditions in (\ref{eq:initCondH}) and (\ref{eq:initCondOH}) are numerically evaluated and presented.  

 \section{System Response Model and Evaluation}
 \label{sec:sysResp}
Solving a system of nonlinear partial differential equations can be a computationally intensive task, especially when considering 3-dimensional (3D) problems. Therefore, in this initial work we consider a 1-dimensional (1D) space and extend the results to higher dimensions in the future. To solve the 1D system of equations in (\ref{eq:reactDiffEqH})--(\ref{eq:reactDiffEqOH}), the finite difference method (FDM) is employed \cite{finiteDiff-book07}. 

In FDM, time and space are discretized into finite intervals. Let $0 \leq t \leq T$ and  $x_a \leq x \leq x_b$ be the time and spatial interval over which a solution is required. In this work, it is assumed that such intervals are divided into equal subintervals of $\Delta t$ and $\Delta x$, respectively.   Let $i$ and $j$ be the index for each subinterval. Then the approximate time corresponding to index $i$ is $t_i = i \Delta t$ and the approximate spatial coordinate for index $j$ is $x_j=x_a+j\Delta x$. Using this scheme, the partial derivatives in (\ref{eq:reactDiffEqH})--(\ref{eq:reactDiffEqOH}) can be approximated as
\begin{align}
	\frac{\partial C}{\partial t}(x_j,t_i) &\approx \frac{C(x_j,t_{i+1})-C(x_j,t_{i})}{\Delta t} \\
	\frac{\partial^2C}{\partial x^2}(x_j,t_i) &\approx  \frac{C(x_{j-1},t_i)-2C(x_{j},t_i)+C(x_{j+1},t_i)}{\Delta x^2} \\
	\frac{\partial C}{\partial x}(x_j,t_i) &\approx  \frac{C(x_{j+1},t_i)-C(x_{j-1},t_i)}{2\Delta x}
\end{align}

Substituting these approximations in (\ref{eq:reactDiffEqH})--(\ref{eq:reactDiffEqOH}), it is possible to find the concentrations at time index $i+1$ from the concentrations at time index $i$ iteratively. The solution for the initial conditions is then obtained by setting the initial concentrations at time index $i=0$ according to (\ref{eq:initCondH}) or (\ref{eq:initCondOH}). In this section, a MATLAB implementation of this technique is used to find a numerical solution of (\ref{eq:reactDiffEqH})--(\ref{eq:reactDiffEqOH}) and hence the system response.

Although analytical solutions for the system model presented in this work do not exist, in the rest of this section, approximate analytical expressions are presented for the case when an acid impulse is transmitted and the case when a base impulse is transmitter. For the acid transmission, it is assumed that no reactions occur in the channel. This can be a valid assumption, if for a single acid impulse the initial concentration of hydrogen ions and hydroxide ions are low compared to the amount of acid released. For the base transmission, it is assumed that the reactions occur only at the receiver and not in the channel. Again, this is a valid assumption if the amount of base released by the transmitter is much larger than the ions in the channel. Note that none of these assumptions hold when considering multiple consecutive transmissions. However, as will be shown in the next section, they could be used to find numerical solutions using FDM for two consecutive transmissions. 

First, let us consider an acid impulse, where the number of acid molecules released is much larger than the concentration of ions in the channel. The approximate response is obtained if it is assumed that there are no chemical interactions in the channel. If the velocity is constant and in the direction from the transmitter to the receiver, and there is a sudden impulse of $N$ moles of particles (i.e. $N\delta(x)$) at time $t=0$, the concentration at the receiver as a function of time is given by 
\begin{align}
	\label{eq:thCons}
	C(t) = N U(t)
\end{align}
with
\begin{align}
\label{eq:Ut}
U(t) = \begin{cases}
\frac{1}{(4\pi D t)^{d/2}}\exp\left(-\frac{(\ell-vt)^2}{4 D t}\right) & t>0 \\
0 & t \leq 0
\end{cases},
\end{align}
where $D$ is the diffusion coefficient of the particle, $d\in\{1,2,3\}$ represents the spatial dimension of the system considered (i.e. 1D, 2D, and 3D), $\ell$ is the separation distance between the transmitter and receiver, and $v$ is the constant velocity. Note that this is the solution of  (\ref{eq:reactDiffEqH}) or (\ref{eq:reactDiffEqOH}) without the reaction terms.  If we assume that there are no chemical reactions (i.e. the hydrogen ions do not interact with any other chemicals in the environment), (\ref{eq:thCons}) plus $10^{-7}$ would represent the concentration of hydrogen ions at the receiver, and its negative log the pH. 

When we consider the reaction as well as the diffusion, analytical expressions do not exist and only numerical evaluation is possible. For this case  (\ref{eq:reactDiffEqH})--(\ref{eq:reactDiffEqOH}) are solved using FDM for the initial condition  (\ref{eq:initCondH}). Note that instead of the Dirac delta function in (\ref{eq:initCondH}), a zero mean Gaussian pulse with a standard deviation of 0.001 is used for numerical evaluations. Let us now compare the approximation with no reactions, with numerical solutions obtained using FDM. In the evaluations, the diffusion coefficient of hydrogen ions is $D_H=9.31\times10^{-5}$ cm$^2$/s \cite{atkinsPhysicalChem09}, and the diffusion coefficient of hydroxide ions is  $D_{OH}=5.03\times10^{-5}$ cm$^2$/s \cite{atkinsPhysicalChem09}. To represent the impulse release of chemicals, it is assumed that 0.001 moles of that chemical are released by the transmitter. 

\begin{figure}[t]
	\begin{center}
		\includegraphics[width=0.70\columnwidth,keepaspectratio]{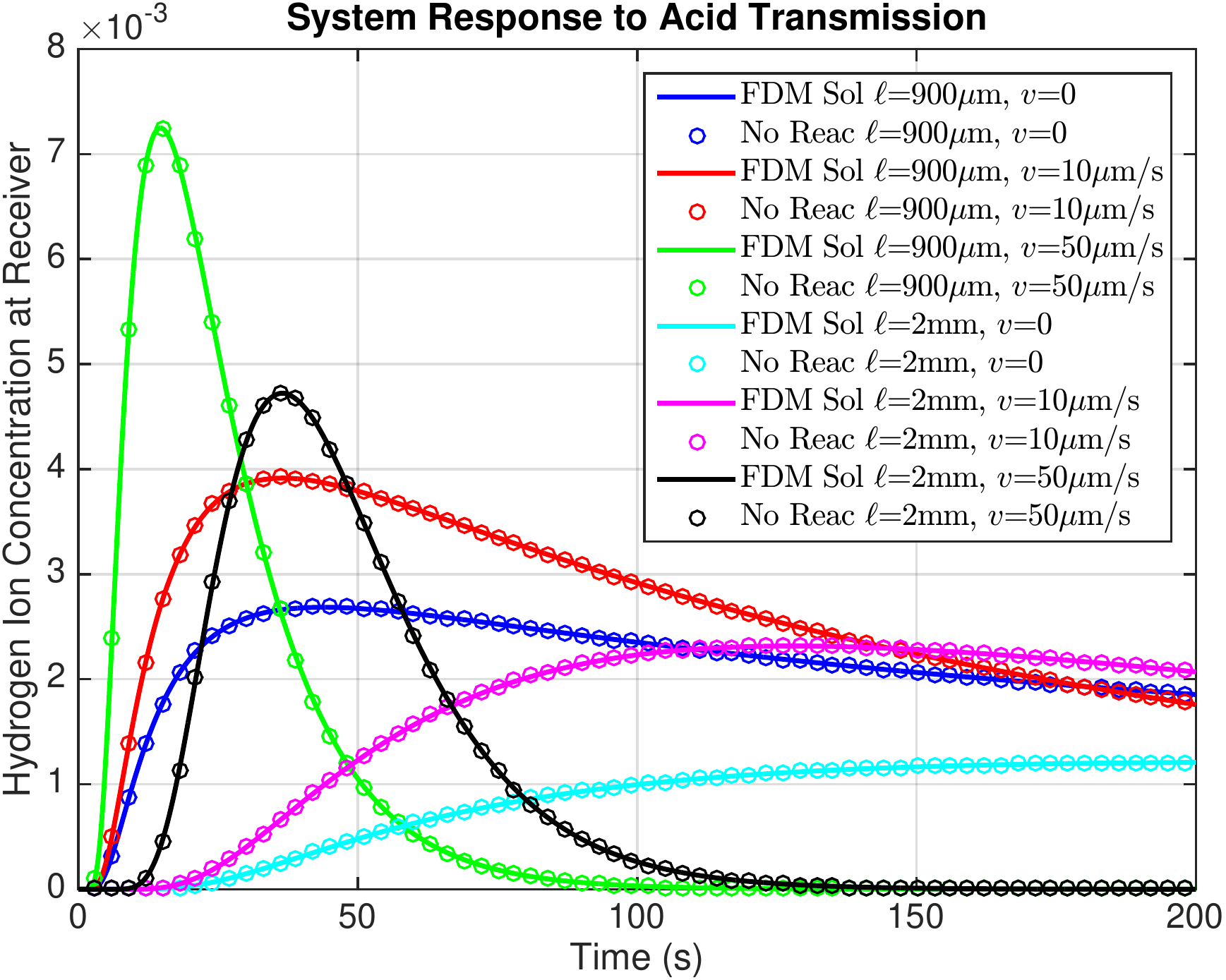}
	\end{center}
	\caption{\label{fig:acidTx} The system response when the transmitter releases a strong acid (i.e. hydrogen ions). The y-axis represent the concentration of hydrogen ions. 
		}
\end{figure}

Figure \ref{fig:acidTx} shows the results for three different flow velocities and two different separation distances: flow velocities of 0, 10 $\mu$m/s, and 50 $\mu$m/s, at 900 $\mu$m and 2 mm. As can be seen in the figure, the system response is the same for the case when the reaction is considered versus the case when no reactions are considered. This is because the initial concentration of hydrogen and hydroxide ions in the channel are much lower than the concentration of the ions released. However, this will not hold when the initial concentration of hydrogen or hydroxide ions in the channel are high.


Let us now consider the case when a strong base (i.e. hydroxide ions) is released by the transmitter. In this case, if the reactions are not considered, then the concentration of hydrogen ions would be constant at $10^{-7}$. 
Therefore, chemical reactions must be included in any approximations. An approximate model in this case can be obtained if it is assumed that the reactions occur only at the receiver. Let $C_{OH}^{(R)}(t)$ be the concentration of hydroxide ions that are released by the transmitter and arrive at the receiver. The $C_{OH}^{(R)}(t)$ can be analytically represented by (\ref{eq:thCons}). 
Since the channel environment is essentially liquid water, (\ref{eq:autoIonEuilb}) must always hold. Let $x$ be the number of water molecules that dissociate into hydrogen and hydroxide ions. Note that in this case, the only source of hydrogen ion is from water autoionization, and $x$ represents the concentration of hydrogen ions. From (\ref{eq:autoIonEuilb}) we have
\begin{align}
\ce{[H+]}\ce{[OH^-]} &=k_w\\
(x)\big(x+C_{OH}^{(R)}(t)\big)&=k_w \\
x^2 + C_{OH}^{(R)}(t)x-k_w&=0.
\end{align}
Solving this quadratic equation for $x$, we obtain an expression for the concentration of hydrogen ions with respect to time:
\begin{align}
\label{eq:consHfromOH}
C_H(t)=& \frac{-C_{OH}^{(R)}(t)+\sqrt{\big(C_{OH}^{(D)}(t)\big)^2+4k_w}}{2}.
\end{align}

\begin{figure}[t]
	\begin{center}
		\includegraphics[width=0.7\columnwidth,keepaspectratio]{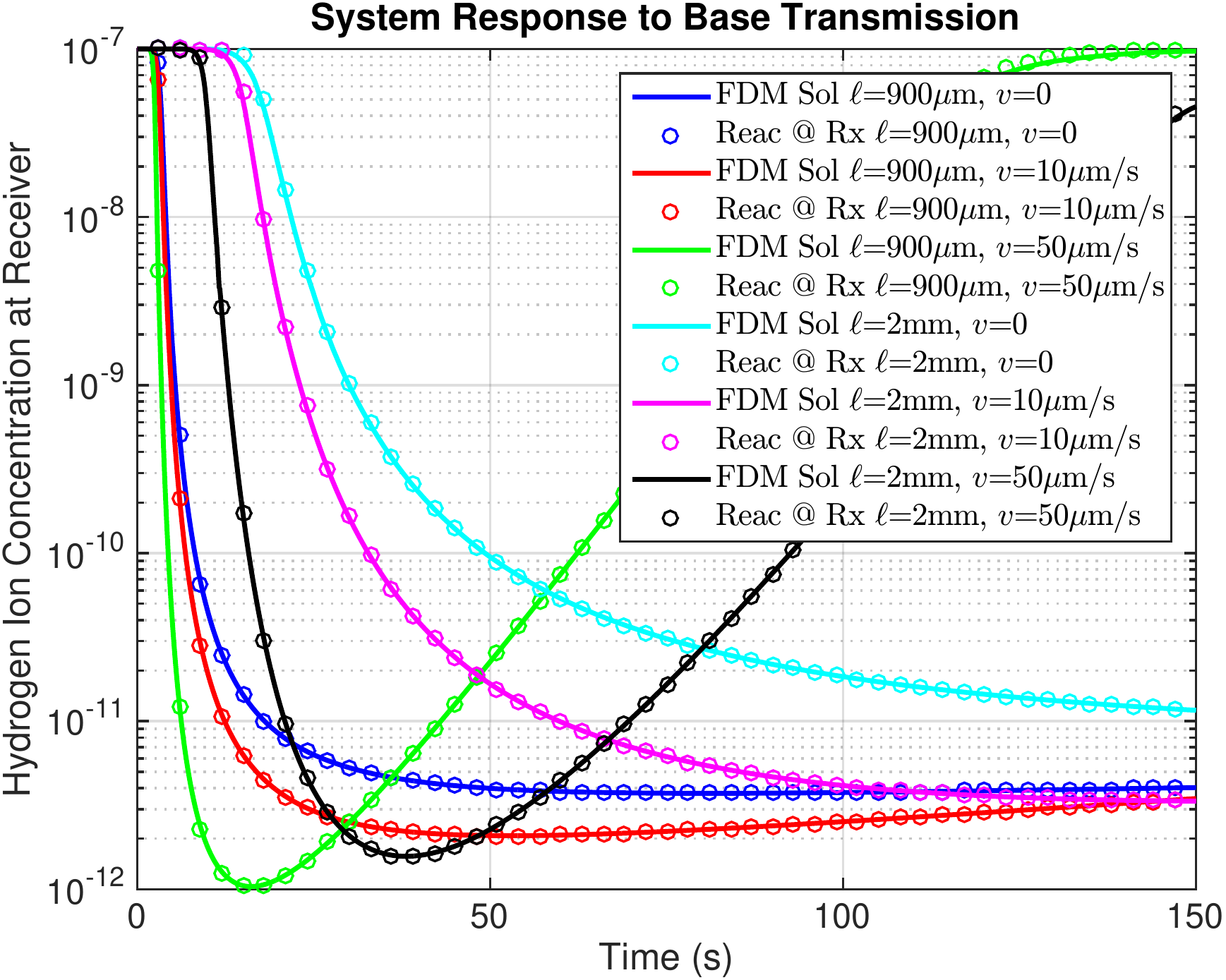}
	\end{center}
	\caption{\label{fig:baseTx} The system response when the transmitter releases a strong base (i.e. hydroxide ions). The y-axis represents the concentration on a log scale. 
		}
\end{figure}

To evaluate the accuracy of this approximation, (\ref{eq:reactDiffEqH})--(\ref{eq:reactDiffEqOH}) are solved for the initial condition (\ref{eq:initCondOH}) using FDM. All the system parameters are the same as the case where a stong acid was released by the transmitter and it is assumed that 0.001 moles of strong base are released. Figure~\ref{fig:baseTx}  shows the results. Unlike Figure \ref{fig:acidTx}, the y-axis in this plot represents the concentration on a log scale. This is to highlight the change in concentration of hydrogen ions which decreases by several orders of magnitude.  As expected, the increasing concentration of hydroxide causes a decrease in concentration of hydrogen ions. The approximation in (\ref{eq:consHfromOH}) (i.e. when it is assumed that the reactions occur only at the receiver) provides a fairly accurate estimate of the concentration of hydrogen ions. This is due to the fact that the initial concentrations of hydrogen and hydroxide ions in the channel are low compared to the concentration released by the transmitter. As will be shown in the next section, this assumption does not hold when consecutive transmissions are considered. 

An important observation is that the concentration of hydrogen ions, and hence the pH level can both increase and decrease depending on whether the transmitter releases an acid or a base. Therefore, more complicated signals can be transmitted to the receiver. For example, as will be shown in the next section, consecutive transmissions could be used to reduce the ISI significantly.

\section{ISI Mitigation Through \\Consecutive Transmissions}
\label{sec:isiMiti}

As was shown in the last section, one of the main benefits of the proposed scheme is the ability of the transmitter to control the increase and decrease of hydrogen ions at the receiver through transmitting acids and bases. Although all concentration-encoded molecular communication techniques can control the increase of concentration at the receiver, there have been no previous works that have reported a transmission technique for decreasing the concentration at the receiver. For example, the enzyme-based techniques proposed in \cite{noe14} rely on the enzymes in the channel, where the transmitter does not have direct control over enzymes. Therefore, one of the most important benefits of the proposed system could be in reducing the ISI. 

As can be seen from Figure~\ref{fig:acidTx} the system response when hydrogen ions (i.e. strong acid) are transmitted by the sender can be very wide, with long tails. One strategy for removing the tail is to transmit hydroxide ions (i.e. strong base) after the acid transmission. Note that for two consecutive transmissions, the simplified approximations that were presented in the previous section (i.e. no reaction assumption, or the assumption that the reactions occur only at the receiver) cannot be applied to the second transmission since the concentration of ions in the channel will be high due to the initial transmission. Therefore, only numerical evaluation is possible in this case. The approximate models from previous section, however, could be used in setting the initial conditions for the second transmission.
\begin{figure}[t]
	\begin{center}
		\includegraphics[width=0.7\columnwidth,keepaspectratio]{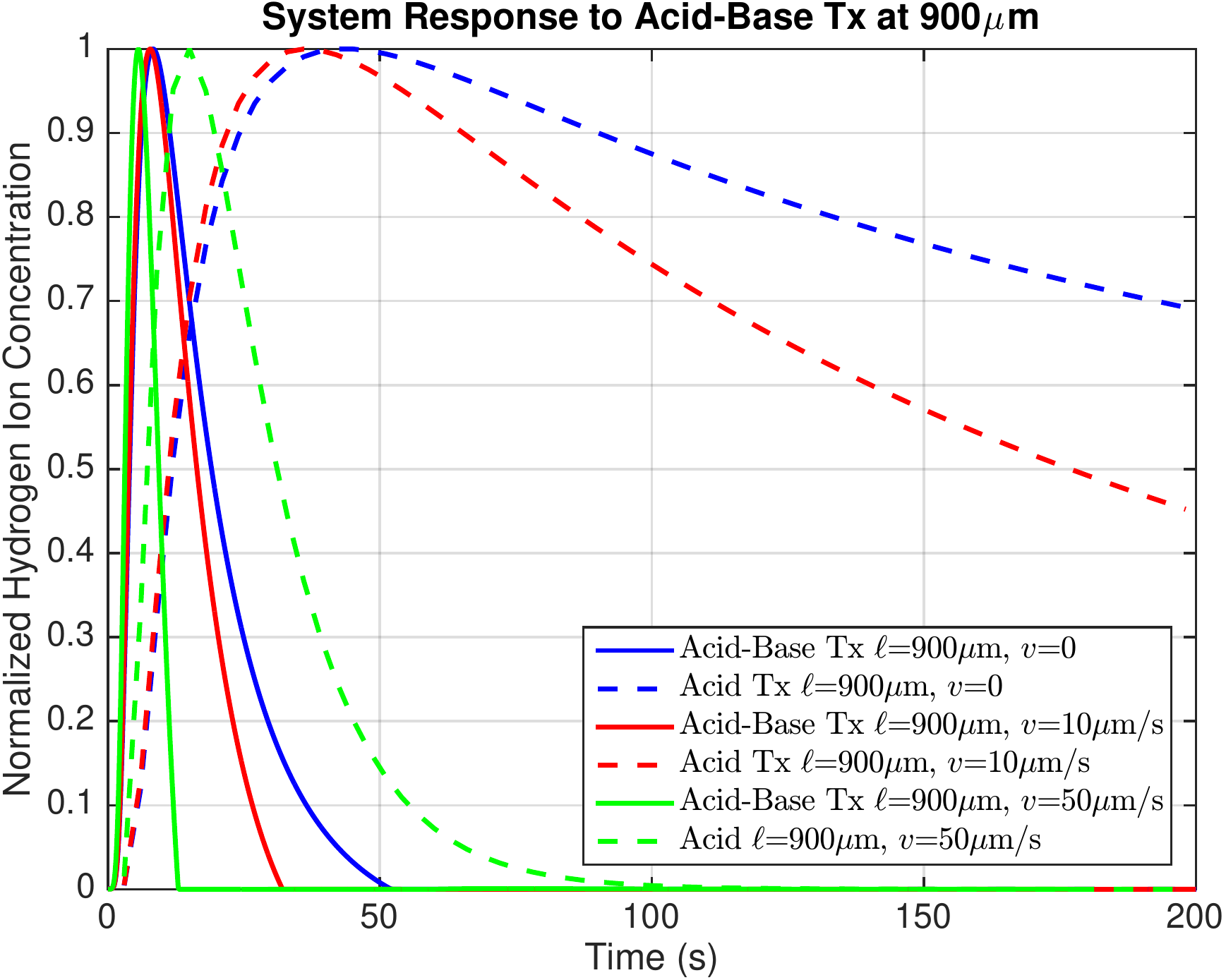}
	\end{center}
	\caption{\label{fig:abTx900} The system response at 900 $\mu$m when the transmitter releases a strong acid only (dashed lines), and when the transmitter releases a strong acid followed by base 0.5 second later (solid lines). 
		}
\end{figure}

In the last section it was shown that (\ref{eq:thCons}) plus $10^{-7}$ is a good approximation for the initial acid transmission. For a constant value of time $T>0$, (\ref{eq:Ut}) as a function of spatial location $\ell$ is a Gaussian probability density function with mean $vT$ and variance $2DT$. Therefore, the initial condition that represents an acid transmission followed by a base transmission $T$ seconds later is given by
\begin{align}
C_{H}(x,t=0)   &=N_{H}\mathcal{N}(vT,2D_aT)+10^{-7}, \\
C_{OH}(x,t=0)&=N_{OH}\delta(x) +\frac{kw}{C_{H}(x,t=0)},
\end{align}
where $\mathcal{N}(\mu,\sigma^2)$ is the Gaussian probability density function with mean $\mu$ and variance $\sigma^2$. For numerical evaluations, instead of the delta function the Gaussian probability density function is used. Note that the same technique could be used to model a base transmission followed by an acid transmission.

\begin{figure}[t]
	\begin{center}
		\includegraphics[width=0.7\columnwidth,keepaspectratio]{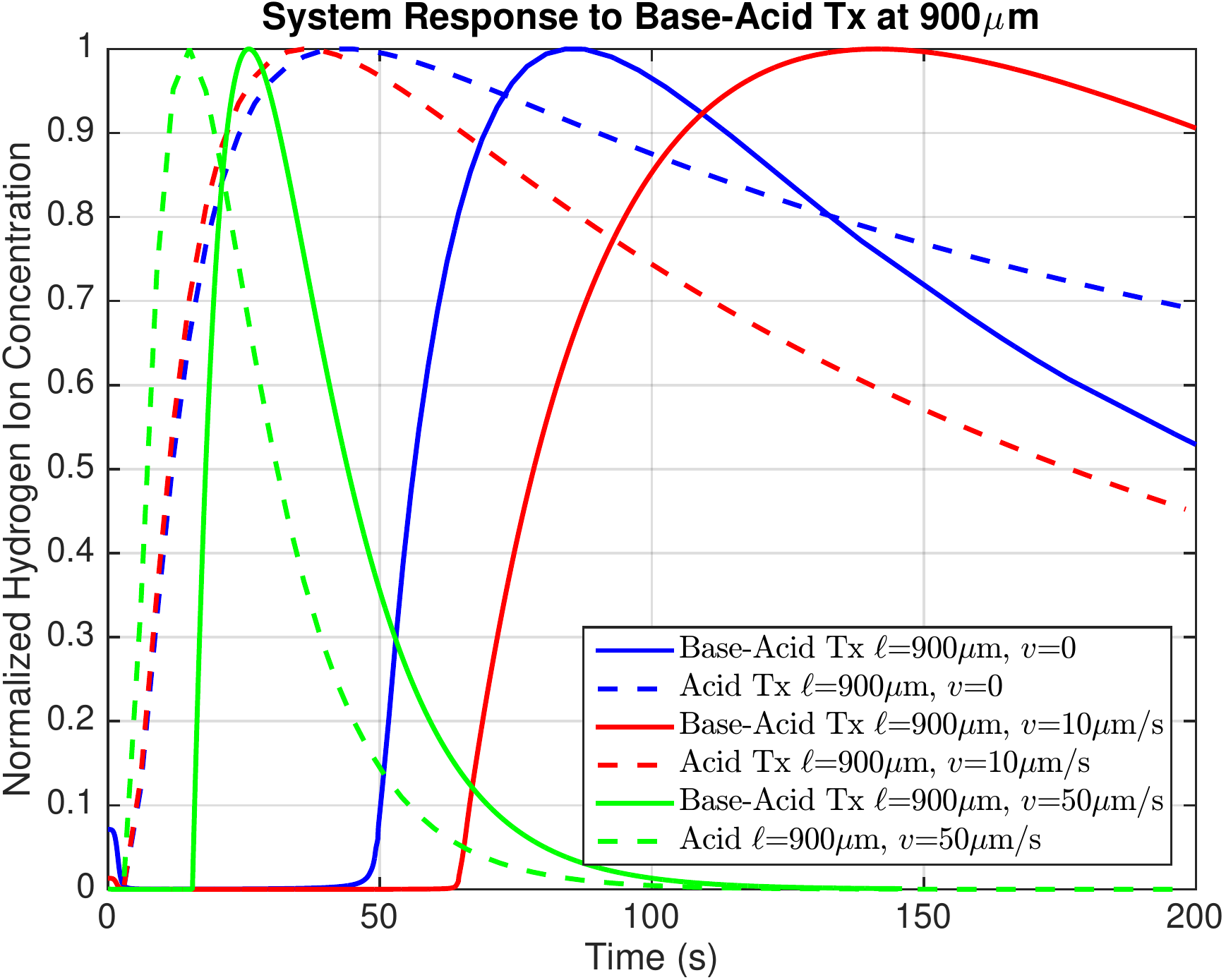}
	\end{center}
	\caption{\label{fig:baTx900} The system response at 900$\mu$m when the transmitter releases a strong acid only (dashed lines), and when the transmitter releases a strong base followed by acid 1 second later (solid lines). 
		}
\end{figure}

First, consider the case when an acid impulse is transmitted followed by a base impulse. Figure~\ref{fig:abTx900} shows the results when the separation distance between the transmitter and the receiver is 900$\mu$m. It is assumed that $T=0.5$ seconds, $N_H = 0.001$ moles, and $N_{OH}=0.001005$ moles. The number of hydroxide ions released is larger to compensate for the smaller diffusion coefficient, which results in a flatter concentration curves. In these plots, the dashed lines represent the case where only an acid impulse is transmitted and the solid lines represent the case where an acid impulse is followed by a base impulse 0.5 seconds later. The plots are normalized by the peak's maximum for easy comparison. As can be seen, the width of the response is decreased significantly and the tails drop quickly toward zero when an acid impulse is followed by a base impulse. Therefore, this can significantly reduce ISI. However, because of the chemical interactions, this reduction in hydrogen ion concentration also means an increase in hydroxide ion concentration, which could effect future transmissions.

To further investigate this effect, the case when a base impulse is transmitted followed by an acid impulse is considered next.  Figure~\ref{fig:baTx900} shows the results when the separation distance between the transmitter and the receiver is 900$\mu$m. It is assumed that $T=1$ seconds, $N_H = 0.001005$ moles, and $N_{OH}=0.001005$ moles. Again the plots are normalized, and the dashed lines represent an acid impulse only and the solid lines represent an acid impulse that follows a base impulse 1 second later. As can be seen in the plots, because of the presence of hydroxide ions in the channel from the initial base impulse, there is a delay in response. However, combining the result from Figures~\ref{fig:abTx900} and \ref{fig:baTx900} it is evident that there is still a significant improvement in ISI. For example, for pure diffusion (blue lines) there is about 50 seconds delay until the pulse drops close to zero in Figures~\ref{fig:abTx900}, and about 50 seconds of delay for the next pulse from Figure~\ref{fig:baTx900}. This combined 100 seconds delay value is much better than the case where only strong acids are transmitted (dashed blue line).

\section{Conclusions and Future Work}
\label{sec:conclusion}
A new and novel concentration-modulated molecular communication scheme using acids, bases and hydrogen ion concentration has been presented. In this scheme, the information can be modulated onto the concentration of hydrogen ions, and this concentration can be controlled through the release of acids and bases at the transmitter. The important benefit of this scheme is that the concentration can be both increased and decreased through transmissions by the sender. This enables inter-symbol interference mitigation schemes as well as the possibility to form a wide array of signal patterns which may be beneficial for control, high-level modulation and multiple access. Another important benefit of this work is the availability of pH sensors at micro-scale and macro-scale for detection, which makes this scheme practical.

As was shown, it is difficult to find analytical expressions for the propagation scheme because of the chemical reactions that result in nonlinear partial differential equations. In the future, we will further study this communication scheme by using stochastic reaction-diffusion simulators, and building an experimental platform. Moreover, we will explore the possibility of forming different signal patterns such as orthogonal signals. Finally, we will consider the case where the transmitter uses a weak acid and a weak base. The weak acids and bases do not dissociate completely in water, and as a result their corresponding model can be more complicated.

\bibliographystyle{IEEEtran}
\bibliography{IEEEabrv,MolCom_YearSorted}

\end{document}